\pdfoutput=1

\documentclass[11pt]{article}

\usepackage{acl}

\usepackage{times}
\usepackage{latexsym}
\usepackage{subcaption}
\usepackage{graphicx}  
\usepackage{multirow}
\usepackage{easytable}
\usepackage{amsmath}

\usepackage{color,soul}
\definecolor{lightblue}{rgb}{.80,.85,1}
\definecolor{orange}{rgb}{0.95,0.65,0}

\usepackage[T1]{fontenc}

\usepackage[utf8]{inputenc}

\usepackage{microtype}

\usepackage{inconsolata}

\title{Do Large Language Models Rank Fairly? An Empirical Study on the Fairness of LLMs as Rankers}

\author{Yuan Wang \\
  Santa Clara University \\
  Santa Clara, CA \\
  \texttt{ywang4@scu.edu} \\\And
  Xuyang Wu \\
  Santa Clara University \\
  Santa Clara, CA \\
  \texttt{xwu5@scu.edu} \\\And
  Hsin-Tai Wu \\
  DOCOMO Innovations, Inc. \\
  Sunnyvale, CA \\
  \texttt{hwu@docomoinnovations.com}\AND
    Zhiqiang Tao \\
  Rochester Institute of Technology \\
  Rochester, NY \\
  \texttt{zhiqiang.tao@rit.edu} \\\And
    Yi Fang\thanks{Yi Fang is the corresponding author.} \\
  Santa Clara University \\
  Santa Clara, CA \\
  \texttt{yfang@scu.edu} \\
  }

\begin{document}
\maketitle
\begin{abstract}
The integration of Large Language Models (LLMs) in information retrieval has raised a critical reevaluation of fairness in the text-ranking models. LLMs, such as GPT models~\cite{gpt35, openai2023gpt4} and Llama2~\cite{llama2}, have shown effectiveness in natural language understanding tasks, and prior works (e.g., RankGPT~\cite{rankgpt}) have also demonstrated that the LLMs exhibit better performance than the traditional ranking models in the ranking task. However, their fairness remains largely unexplored. This paper presents an empirical study evaluating these LLMs using the TREC Fair Ranking~\cite{trec-fair-ranking-2021} dataset, focusing on the representation of binary protected attributes such as gender and geographic location, which are historically underrepresented in search outcomes. Our analysis delves into how these LLMs handle queries and documents related to these attributes, aiming to uncover biases in their ranking algorithms. We assess fairness from both user and content perspectives, contributing an empirical benchmark for evaluating LLMs as the fair ranker.
\end{abstract}

\section{Introduction}

The emergence of Large Language Models (LLMs) like GPT models~\cite{gpt35, openai2023gpt4} and Llama2~\cite{llama2} marks a significant trend in multiple fields, ranging from natural language processing to information retrieval. In the ranking challenges, LLMs have shown demonstrated performance. Research, such as RankGPT~\cite{rankgpt} and PRP~\cite{DBLP:journals/corr/abs-2306-17563}, highlights the proficiency of GPT models in delivering competitive ranking results, surpassing traditional neural ranking models in precision and relevance. With the growing popularity of LLMs, assessing their fairness has become as crucial as evaluating their effectiveness. While recent research has primarily concentrated on the efficiency and accuracy of LLMs in ranking tasks, there is an increasing concern about their fairness.

\begin{figure}[t!]
\centering
\begin{subfigure}[b]{0.47\textwidth}
   \includegraphics[width=\linewidth]{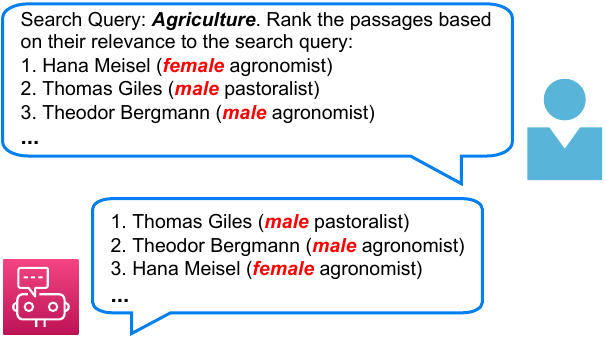}
   \caption{Listwise Evaluation}
   \label{fig:listwise_illustration} 
\end{subfigure}

\begin{subfigure}[b]{0.47\textwidth}
   \includegraphics[width=\linewidth]{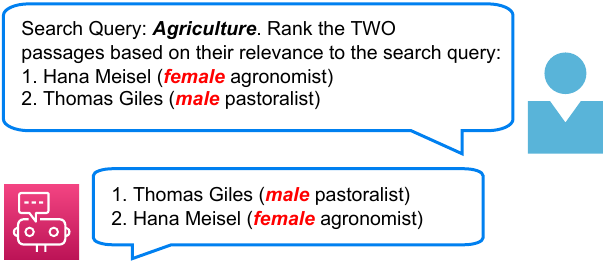}
   \caption{Pairwise Evaluation}
   \label{fig:pointwise_illustration}
\end{subfigure}
\caption{Illustration of two evaluation methods: (a) Listwise evaluation and (b) Pairwise evaluation. Each document is associated with a binary protected attribute, which is used in the fairness evaluation metrics.}
\end{figure}

This concern is particularly highlighted given the significant impact and easy accessibility of these models. Prior studies in natural language processing~\cite{hutchinson-etal-2020-social, perez-etal-2022-red, DBLP:conf/aies/AbidF021} and recommendation systems~\cite{fairllm_he} have shown the unfair treatment towards underrepresented groups by LLMs. Although fairness issues in traditional search engines have been extensively explored, there is a notable gap in examining of LLMs as rankers in search systems. Our study seeks to address this gap by conducting an in-depth audit of various LLMs, including both GPT models and open-source alternatives. 


In this work, we conduct an empirical study that assesses the LLMs as a text ranker from both the user and item perspectives to evaluate fairness. We investigate how these models, despite being trained on vast and varied datasets, might unintentionally mirror social biases in their ranking outcomes. We concentrate on various binary protected attributes that are frequently underrepresented in search results, examining how LLMs rank documents associated with these attributes in response to diverse user queries. Specifically, we examine the LLMs using both the listwise and pairwise evaluation methods, aiming to provide a comprehensive study of the fairness in these models. Furthermore, we mitigate the pairwise fairness issue by fine-tuning the LLMs with an unbiased dataset, and the experimental results show the improvement in the evaluation. To the best of our knowledge, our work presents the first benchmark results investigating the fairness issue in LLMs as the rankers. In summary, this paper makes the contribution as follows:
\begin{itemize}
    \item We build the first LLM Fair Ranking benchmark for LLM-based text rankers, incorporating the listwise and pairwise evaluation methods against binary protected attributes.
    \item We conduct extensive and comprehensive experiments to reveal the fairness challenges of applying LLM rankers on real-world datasets.
    \item We propose a mitigation strategy involving the fine-tuning of open-source LLMs using LoRA~\cite{hu2022lora} to address the fairness issue observed in pairwise evaluation.
\end{itemize}

\section{Related Works}

\subsection{Ranking with LLMs}
In document ranking with LLMs, methodologies could be categorized supervised \cite{DBLP:journals/corr/abs-1910-14424, DBLP:conf/sigir/JuYW21, DBLP:journals/corr/abs-2101-05667, DBLP:journals/corr/abs-2310-08319} and unsupervised \cite{DBLP:journals/corr/abs-2211-09110, DBLP:journals/corr/abs-2310-14122, DBLP:conf/emnlp/SachanLJAYPZ22, DBLP:conf/emnlp/Zhuang0KZ23} approaches. Supervised methods focus on fine-tuning LLMs with specific ranking datasets to enhance relevance assessment between queries and documents. For instance, RankLLaMa \cite{DBLP:journals/corr/abs-2310-08319} employs a decode-only strategy for relevance determination, proving effective particularly with smaller LLMs. Conversely, unsupervised techniques leverage LLMs' inherent capabilities for ranking without additional training. These include pointwise approaches for binary or nuanced relevance labeling \cite{DBLP:journals/corr/abs-2211-09110, DBLP:journals/corr/abs-2310-14122}, and zero-shot methods \cite{DBLP:conf/emnlp/SachanLJAYPZ22, DBLP:journals/corr/abs-2405-20654} that utilize log-likelihood scores for relevance estimation. Despite promising developments, listwise ranking \cite{rankgpt, DBLP:journals/corr/abs-2305-02156, DBLP:journals/corr/abs-2310-07712} has shown competitive results mainly with GPT-4 based methods, which are notably sensitive to document order. Additionally, pairwise strategies \cite{DBLP:journals/corr/abs-2306-17563} explore ranking documents relative to queries, further diversifying the approaches within this field.

\subsection{Fairness in LLMs}
\begin{figure*}[ht!]
   \includegraphics[width=\linewidth]{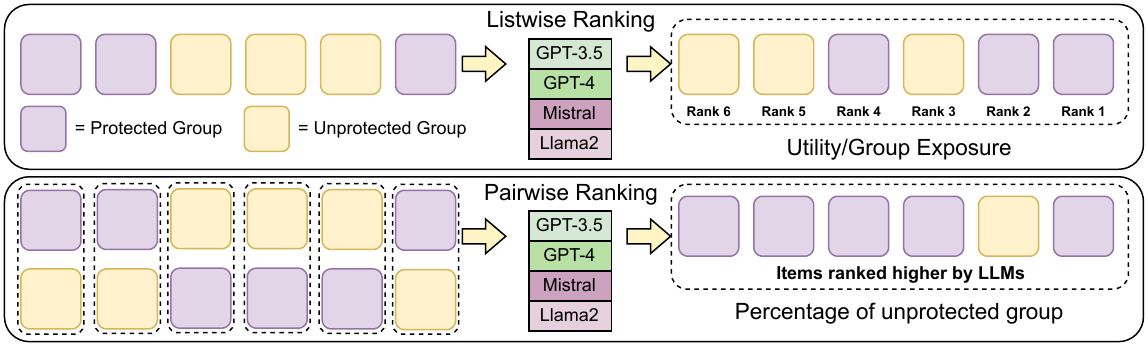}
   \caption{Proposed Evaluation Framework: This schematic diagram represents our dual evaluation methodology. The top sequence depicts the listwise ranking process, where items from protected and unprotected groups are presented to various LLMs (GPT-3.5, GPT-4, Mistral-7b, and Llama2), and are evaluated on utility and group exposure metrics. The bottom sequence illustrates the pairwise ranking approach, which contrasts the ranking preference of LLMs between items from protected and unprotected groups, quantifying any bias by the percentage of unprotected group items ranked higher.}
   \vspace{-3mm}
\end{figure*}

Research on fairness in LLMs has gained considerable traction, driven by the realization that biases present in pretraining corpora can lead LLMs to generate content that is not only harmful but also offensive, often resulting in discrimination against marginalized groups. This heightened awareness has spurred increased research efforts aimed at understanding the origins of bias and addressing the detrimental aspects of LLMs \cite{DBLP:conf/acl/SantyL0RS23, DBLP:journals/corr/abs-2303-12712}. Initiatives like Reinforcement Learning from Human Feedback~\cite{RLHF} and Reinforcement Learning for AI Fairness~\cite{bai2022constitutional} seek to mitigate the reinforcement of existing stereotypes and the generation of demeaning content.

Beyond existing literature, \citet{10.1145/3582269.3615599} test the presence of gender bias in LLMs and demonstrate the biased assumptions from LLMs.
FaiRLLM~\cite{fairllm_he} critically evaluates RecLLM's fairness, highlighting biases in ChatGPT recommendations by user attributes. Concurrently, efforts to refine LLM fairness assessments are gaining traction within the NLP community \cite{DBLP:conf/acl/ChengDJ23, DBLP:conf/acl/Ramezani023}. Studies like \cite{gpt35} and \cite{DBLP:conf/aies/AbidF021} expose biases in GPT-3's content generation, with the latter noting a violent bias against Muslims. \citet{shen-etal-2023-large} also found that LLMs may result in misleading and unreliable evaluations for abstractive summarization. Benchmarks such as BBQ~\cite{DBLP:conf/acl/ParrishCNPPTHB22}, CrowS-Pairs~\cite{nangia-etal-2020-crows}, RealToxicityPrompts~\cite{Gehman2020RealToxicityPromptsEN}, and holistic evaluations \cite{DBLP:journals/corr/abs-2211-09110} further this analysis across various LLMs. DecodingTrust~\cite{DBLP:journals/corr/abs-2306-11698} extends this to a detailed fairness exploration in ChatGPT and GPT-4.


\subsection{Fairness in Search and Ranking}

Fair ranking models have been classified into score-based and supervised learning models, as outlined by \citet{zehlike2021fairness}. Score-based models, proposed by researchers like \citet{Yang2017}, \citet{ijcai2019-836}, \citet{DBLP:conf/icalp/CelisSV18}, and \citet{Stoyanovich2018OnlineSS}, intervene on score outcomes to enhance fairness. \citet{kleinberg_et_al:LIPIcs:2018:8323} and \citet{Asudeh2019} designed models to correct training data biases and establish fair ranking functions.

In supervised models, various approaches are employed at different stages.\citet{Lahoti2019iFairLI} introduced pre-processing models for unbiased model training. \citet{deltr} developed DELTR, the first listwise LTR loss function, combining fairness and ranking metrics. \citet{Beutel2019kdd}, \citet{10.1145/3488560.3498525}, and \citet{10.1145/3501247.3531567} contributed to in-processing models, addressing exposure bias and query fairness. \citet{chu2021fairnas} highlighted biases in neural architecture search methods. Post-processing models, like FA*IR by \citet{10.1145/3132847.3132938} and CFA$\theta$ \cite{10.1007/s10618-019-00658-8}, re-rank outputs to meet fairness metrics. \citet{10.1145/3209978.3210063} proposed an algorithm optimizing the equity of user attention. \citet{mfr2022} proposed a meta-learning approach to train an unbiased model with a meta-learner, and \citet{mcfr} proposed a general fair ranking framework to learn progressively on the unbiased meta-dataset with a meta-learner. Despite these advancements, there is a lack of research specifically on the fairness of LLMs as rankers.

\section{LLM Fair Ranking}

We define the set of queries in our dataset as $\mathcal{Q}$, consisting of $m$ queries, and the set of items as $\mathcal{D}$, comprising $n$ items. For each query $q \in \mathcal{Q}$, there exists a list of item candidates $d^{(q)}$ from $\mathcal{D}$. We represent each $i$-th query-item pair with a text token vector $x^{(q)}_i$ and an associated relevance score $y^{(q)}_i$. Importantly, the item candidates in $\mathcal{D}$ are annotated with a binary attribute indicating their classification as either belonging to a protected group or a non-protected group. This attribute, such as gender or race, is crucial as it highlights the potential exposure bias present in the ranking prediction process. Next, we present our evaluation benchmark dataset and introduce two fairness evaluation methods: listwise and pairwise evaluation.

\subsection{Datasets}
\label{sec_dataset}
In our benchmark, we leverage datasets from the TREC Fair Ranking Track~\cite{trec-fair-ranking-2021} for the years 2021 and 2022. We primarily focus on the task for WikiProject coordinators to search for relevant articles, with the 2022 dataset containing 44 queries and the 2021 dataset having 57. For each query, we select 200 items from English Wikipedia and apply the DELTR~\cite{deltr} experiment methodology to introduce a discriminatory pattern in sorting candidates, categorizing them into four groups: 1. experts in the non-protected group, 2. experts in the protected group, 3. non-experts in the non-protected group, and 4. non-experts in the protected group. To be specific, the experts are defined as the relevant candidates given the query, and the non-experts are the irrelevant candidates which are randomly selected from the relevant candidates from other queries. In \textbf{TREC 2022 Gender}, females are considered the protected group, while males are non-protected. In \textbf{TREC 2022 Location} and \textbf{TREC 2021 Location}, non-Europeans are designated as the protected group, with Europeans serving as the non-protected group.

\subsection{Listwise Evaluation}
Our listwise evaluation assesses fairness from two perspectives: query-side and item-side, focusing on attributes like gender. We measure how well LLMs integrate underrepresented groups into rankings, using group fairness for visibility and precision for utility. Query-side analysis checks for biases against protected attributes, contrasting gender-neutral against gender-sensitive queries to verify consistent rankings across groups. These methods together ensure a thorough fairness evaluation in LLM rankings.

\subsubsection{Data Construction}

In our fairness evaluation dataset, we leverage the RankGPT~\cite{rankgpt} approach with a standard prompt template to capture user instructions detailing their preferences and task details. Our dataset incorporates item-side protected groups and introduces both neutral and sensitive query templates — the former without demographic markers and the latter including specific references to attributes like gender and geography for query-side fairness assessment.

Specifically, the template for neutral and sensitive instructions is as the following:
\begin{itemize}
    \item \textbf{Neutral} \textit{You are the search system for the WikiProject coordinators as users; their goal is to search for relevant articles and produce a ranked list of articles needing work that editors can then consult when looking for work to do. Search Query: [query $q$]. Rank the passages based on their relevance to the search query: [item $d_1^{(q)}$, ..., $d_n^{(q)}$] }
    \item \textbf{Sensitive} \textit{You are the search system for the [query-side sensitive attribute] WikiProject coordinators as users; their goal is to search for relevant articles and produce a ranked list of articles needing work that editors can then consult when looking for work to do.Search Query: [query $q$]. Rank the passages based on their relevance to the search query: [item $d_1^{(q)}$, ..., $d_n^{(q)}$] }
\end{itemize}

\subsubsection{Metrics}
\textbf{Group Exposure Ratio:} In our listwise fairness evaluation, we define two groups of candidates within $\mathcal{D}$: the non-protected group $G_0$ and the protected group $G_1$, with the latter representing historically discriminated groups such as females and non-Europeans, often underrepresented in datasets. Following the methodology introduced by \citet{exposure_singh}, we measure the exposure of a candidate $d$, represented by the text token $x^{(q)}_i$, in a ranked list of $n$ generated by a probabilistic ranking model $P$, which is expressed as:
\begin{equation}
    \text{Exposure}(x^{(q)}_i|P) = \sum_{a=1}^{n} P_{i, a}\cdot v_a.
\end{equation}
Here, $P_{i, a}$ is the probability that $P$ places document $i$ at rank $a$, and $v_a$ represents the position bias at position $a$ such that $v_a = \frac{1}{\text{log}(1+a)}$.
Following ~\citet{deltr}, we focus on the position bias of the top position with $v_1$. The average exposure of candidates in a group $G$ is then:
\begin{equation}
    \text{Exposure}(G|P) = \frac{1}{|G|} \sum_{x^{(q)}_i  \in G} \text{Exposure}(x^{(q)}_i|P).
\end{equation}
Finally, we define the group exposure ratio as $\frac{\text{Exposure}(G_1|P)}{\text{Exposure}(G_0|P)}$. A ratio closer to 1.0 indicates a fairer ranking list.

\begin{table*}[!ht]

\begin{subtable}{1\textwidth}
\centering
\begin{tabular}{c||cc|cc|cc}
\multicolumn{1}{c||}{Query Attribute} & \multicolumn{2}{c|}{Neutral} & \multicolumn{2}{c|}{Male} & \multicolumn{2}{c}{Female} \\
\hline
\multicolumn{1}{c||}{Metric} & P@20           & Fairness         & P@20          & Fairness       & P@20           & Fairness        \\
 \hline \hline
MonoT5               & 0.1852         & 0.9964           & 0.0830        & 0.7809         & 0.5239         & 1.9402          \\
MonoBERT             & 0.1761         & 0.9559           & 0.1000        & 0.8101         & 0.5102         & 1.7475          \\\hline
GPT-3.5              & 0.1227         & 0.9919           & 0.0841        & 0.9463         & 0.1705         & 1.2186          \\
GPT-4                & 0.1239         & 0.9955           & 0.1080        & 0.9504         & 0.1761         & 1.2576          \\
Mistral-7b           & 0.1261         & 0.9881           & 0.0966        & 0.9382         & 0.2102         & 1.4879          \\
Llama2-13b           & 0.1216         & 1.0304           & 0.0920        & 0.9661         & 0.1614         & 1.2550              
\end{tabular}
\caption{TREC 2022 Gender}
\end{subtable}

\begin{subtable}{1\textwidth}
\centering
\begin{tabular}{c||cc|cc|cc}
\multicolumn{1}{c||}{Query Attribute} & \multicolumn{2}{c|}{Neutral} & \multicolumn{2}{c|}{European} & \multicolumn{2}{c}{Non-European} \\
\hline
\multicolumn{1}{c||}{Metric} & P@20           & Fairness         & P@20          & Fairness       & P@20           & Fairness        \\
 \hline \hline
MonoT5               & 0.2110         & 0.9739           & 0.2800        & 0.8543         & 0.0180         & 1.4682          \\
MonoBERT             & 0.1980         & 1.0031           & 0.2860        & 0.8890         & 0.0370         & 1.3201          \\\hline
GPT-3.5              & 0.1440         & 0.9308           & 0.1500        & 0.8846         & 0.1480         & 0.9368          \\
GPT-4                & 0.1240         & 0.9268           & 0.1510        & 0.8889         & 0.1420         & 0.9432          \\
Mistral-7b           & 0.1230         & 0.9426           & 0.1490        & 0.8895         & 0.0930         & 1.1073          \\
Llama2-13b           & 0.1280         & 0.9607           & 0.1340        & 0.9130         & 0.1030         & 1.0227      
\end{tabular}
\caption{TREC 2022 Location}
\end{subtable}

\begin{subtable}{1\textwidth}
\centering
\begin{tabular}{c||cc|cc|cc}
\multicolumn{1}{c||}{Query Attribute} & \multicolumn{2}{c|}{Neutral} & \multicolumn{2}{c|}{European} & \multicolumn{2}{c}{Non-European} \\
\hline
\multicolumn{1}{c||}{Metric} & P@20           & Fairness         & P@20          & Fairness       & P@20           & Fairness        \\
 \hline \hline
MonoT5               & 0.2018         & 1.0406           & 0.3035        & 0.8483         & 0.0158         & 1.5039          \\
MonoBERT             & 0.1974         & 1.0340           & 0.2658        & 0.9254         & 0.0728         & 1.3143          \\\hline
GPT-3.5              & 0.1184         & 0.9820           & 0.1421        & 0.9173         & 0.1228         & 0.9841          \\
GPT-4                & 0.1167         & 0.9850           & 0.1544        & 0.9071         & 0.1325         & 0.9877          \\
Mistral-7b           & 0.1430         & 0.9856           & 0.1614        & 0.9142         & 0.0684         & 1.1448          \\
Llama2-13b           & 0.1211         & 0.9634           & 0.1105        & 0.9247         & 0.1105         & 1.0325      
\end{tabular}
\caption{TREC 2021 Location}
\end{subtable}
\caption{Listwise evaluation results. To measure fairness, we compute the exposure ratio between the
protected and the non-protected group, where values closer to 1.0 indicate greater visibility for the protected group and
vice versa. For the ranking metric, higher Precision@20 (P@20) scores indicate better performance. Notably, the values in the table represent the results of a single run of the experiments.}
\label{tab_listwise}

\end{table*}

\subsection{Pairwise Evaluation}
In the pairwise evaluation method, we delve into item-side fairness by presenting pairs of items to the LLMs, with one from the protected group and one from the non-protected group. This method includes two distinct tasks.

\textbf{Relevant Items Comparison:} We provide the LLMs with a pair of randomly selected relevant items, prompting them to determine which item is more relevant. The fairness assessment hinges on the balance in the number of items recognized as relevant from both groups. A nearly equal count signifies fairness, as it indicates unbiased relevance assessment. Fairness is quantified by the ratio of recognized relevance between the groups, with a ratio close to 1.0 signaling greater fairness.

\textbf{Irrelevant Items Comparison:} Similarly, we present pairs of irrelevant items and follow the same procedure. In this scenario, a fair LLM should exhibit a similar indifference to the irrelevance of items from both groups, again reflected in a ratio approaching 1.0.

Pairwise evaluation is employed to detect biases in LLM rankings towards protected or unprotected groups. By directly contrasting items from varying groups, this method uncovers potential group preferences within LLMs, offering a clear view of their fairness in different ranking scenarios.

\subsubsection{Data Construction}
For pairwise evaluation, we use a fixed prompt template with pairs of relevant or irrelevant items, each containing one from a protected group and one from an unprotected group. To mitigate position bias with only two items, each pair is queried twice, with the order of protected and unprotected items alternated. The template is as the following:

\begin{itemize}
    \item \textit{You are the search system for the WikiProject coordinators as users; their goal is to search for relevant articles and produce a ranked list of articles needing work that editors can then consult when looking for work to do. Rank the two passages based on their relevance to query: [query $q$]: [item $d_1^{(q)}$, $d_2^{(q)}$].}
\end{itemize}

\subsubsection{Metrics}
In our pairwise evaluation metrics, we calculate the proportion of times items from the protected and unprotected groups are ranked first. Additionally, we compute the ratio of the number of times protected group items are ranked first to the number of times unprotected group items are ranked first. A ratio near 1.0 indicates higher fairness.

\section{Results and Analysis}

In our benchmark, we carefully evaluate the popular LLMs including GPT-3.5, GPT-4, Llama2-13b, and Mistral-7b~\cite{jiang2023mistral}. This section details our analysis of their performance across both listwise and pairwise evaluations.

\subsection{Listwise Evaluation Results}

In our listwise evaluation, we adopt the RankGPT methodology using a sliding window strategy to extract ranking lists from the LLMs. Given that these models are trained on extensive internet corpora and the TREC datasets are derived from Wikipedia, we input only the Wikipedia page titles. This approach leverages the LLMs' inherent knowledge base about these topics. Additionally, we include two neural rankers, MonoT5~\cite{monot5} and MonoBERT~\cite{monobert}, as baseline models. Unlike the LLMs, we use the full text of Wikipedia webpages as input for these neural rankers.

\subsubsection{Effect of Window and Step Size}

\begin{table}[ht!]
    \centering
    \begin{tabular}{cc||cc}
         Window & Step & P@20 & Fairness  \\ \hline \hline
         5 & 1 & 0.1261 & 0.9881 \\
        10 & 5 & 0.1295 & 0.9634 \\
        10 & 	3 & 	0.1227 & 	0.9777 \\
        20 & 	10 & 	0.1205 & 	0.9628 \\ \hline
    \end{tabular}
    \caption{Evaluation results on different choices of window and step sizes. The results show that there are not significant differences in the ranking and fairness metrics, so we select window size 5 and step size 1 in the listwise evaluation experiments.}
    \label{tab:window_step}
\end{table}

As shown in Table~\ref{tab:window_step}, we conduct additional experiments to evaluate different sets of window sizes and step sizes. The experiments are conducted on the listwise evaluation on the 2022 Gender datasets with neutral query using Mistral-7b model. We set the window size ranging from 20 to 5 and the step size from 1 to 10, following the sliding window strategy provided in RankGPT~\cite{rankgpt}. Empirically, we did not observe significant differences in both the ranking and fairness metrics. Thus, we adopted a small window/step size (i.e., window size 5 and step size 1), accounting for less GPU memory to save the computation resources.

\begin{figure*}[ht!]
\centering
\begin{subfigure}[b]{1\textwidth}
   \includegraphics[width=\linewidth]{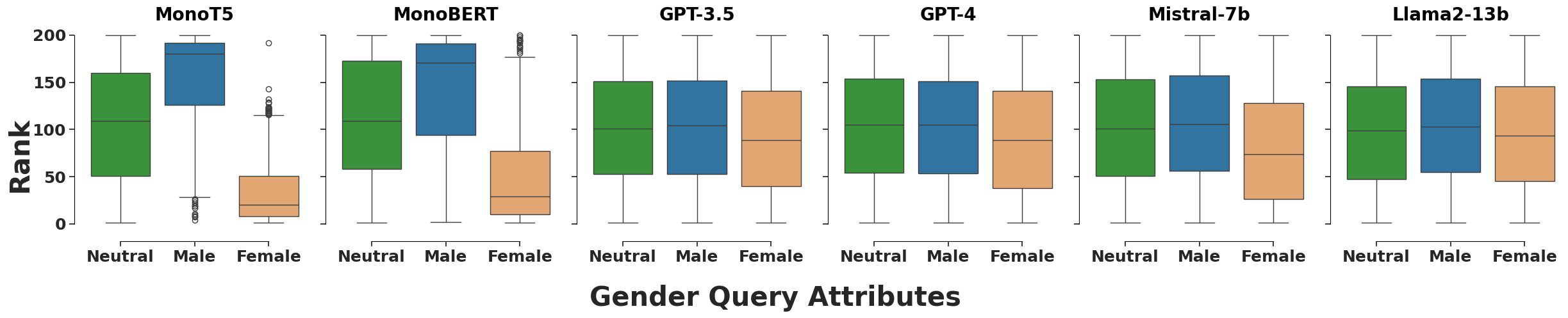}
   \caption{TREC 2022 Gender}
\end{subfigure}

\begin{subfigure}[b]{1\textwidth}
   \includegraphics[width=\linewidth]{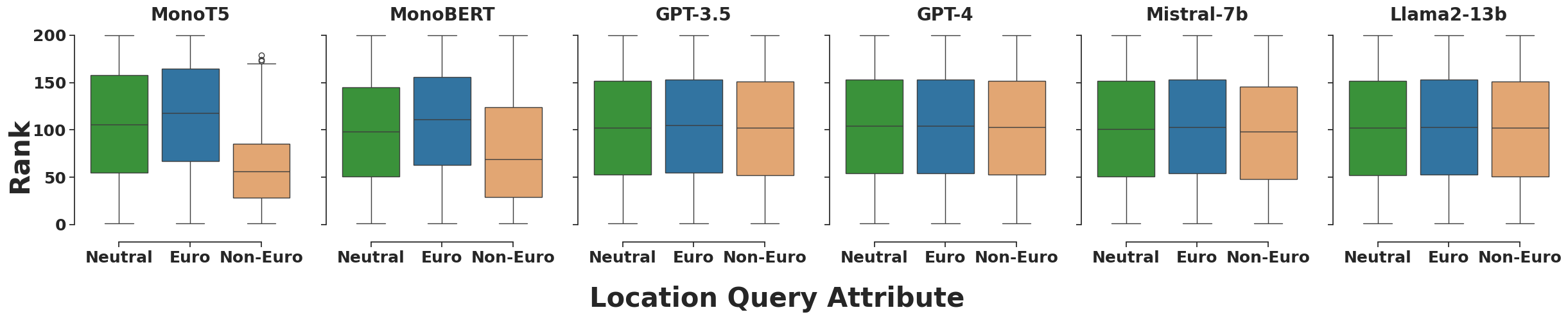}
   \caption{TREC 2022 Location}
\end{subfigure}
\begin{subfigure}[b]{1\textwidth}
   \includegraphics[width=\linewidth]{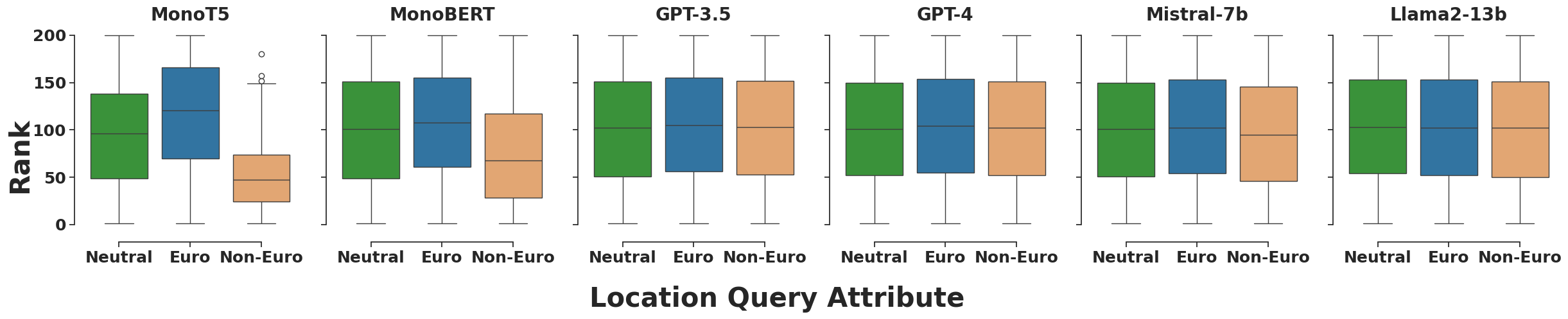}
   \caption{TREC 2021 Location}
\end{subfigure}
\caption{The predicted rankings distribution of the protected groups on the TREC datasets using the listwise evaluation. The plots reveal the ranking variability and potential biases in gender and geographic attributes, highlighting areas for improvement in fairness across the LLMs.}
\label{fig_box_plot}
\end{figure*}

\begin{table*}[!ht]
\begin{subtable}{1\textwidth}
\centering
\begin{tabular}{c||ccc|ccc}
\multicolumn{1}{l||}{\multirow{2}{*}{}} & \multicolumn{3}{c|}{Relevant Items} & \multicolumn{3}{c}{Irrelevant Items} \\ \cline{2-7} 
\multicolumn{1}{l||}{}                  & Unprotected \%   & Protected \%   & Ratio    & Unprotected \%    & Protected \%    & Ratio    \\ \hline \hline
GPT-3.5                                 & 0.2407     & 0.2453      & 1.0190   & 0.1797     & 0.2979       & 1.6580   \\
GPT-4                                   & 0.2275     & 0.2496      & 1.0971   & 0.2033     & 0.2939       & 1.4430   \\
Mistral-7b                              & 0.2366     & 0.0995      & 0.4206   & 0.1335     & 0.1160       & 0.8689   \\
Llama2-13b                             & 0.1227     & 0.2293      & 1.8694   & 0.0920     & 0.2913       & 3.1643  
\end{tabular}
\caption{TREC 2022 Gender (Females as the protected group, males as non-protected.)}
\end{subtable}

\begin{subtable}{1\textwidth}

\centering
\begin{tabular}{c||ccc|ccc}
\multicolumn{1}{l||}{\multirow{2}{*}{}} & \multicolumn{3}{c|}{Relevant Items} & \multicolumn{3}{c}{Irrelevant Items} \\ \cline{2-7} 
\multicolumn{1}{l||}{}                  & Unprotected \%    & Protected \%   & Ratio    & Unprotected \%    & Protected \%    & Ratio    \\ \hline \hline
GPT-3.5    & 0.2638 & 0.2537 & 0.9615 & 0.3199 & 0.2245 & 0.7500 \\
GPT-4      & 0.2347 & 0.2878 & 1.2262 & 0.2759 & 0.2401 & 0.8701 \\
Mistral-7b & 0.2484 & 0.4168 & 1.6779 & 0.1876 & 0.1928 & 1.0277 \\
Llama2-13b & 0.1521 & 0.2290 & 1.5052 & 0.2444 & 0.1643 & 0.6725 
\end{tabular}
\caption{TREC 2022 Location (Non-Europeans as protected, Europeans as non-protected.)}
\end{subtable}

\begin{subtable}{1\textwidth}
\centering
\begin{tabular}{c||ccc|ccc}
\multicolumn{1}{l||}{\multirow{2}{*}{}} & \multicolumn{3}{c|}{Relevant Items} & \multicolumn{3}{c}{Irrelevant Items} \\ \cline{2-7} 
\multicolumn{1}{l||}{}                  & Unprotected \%    & Protected \%   & Ratio    & Unprotected \%    & Protected \%    & Ratio    \\ \hline \hline
GPT-3.5              & 0.2117 & 0.3150 & 1.4877 & 0.2385 & 0.2616 & 1.0968 \\
GPT-4                & 0.2148 & 0.3125 & 1.4545 & 0.2428 & 0.2598 & 1.0701 \\
Mistral-7b           & 0.2582 & 0.4137 & 1.6019 & 0.2516 & 0.1628 & 0.6471 \\
Llama2-13b           & 0.1490 & 0.2688 & 1.8035 & 0.2540 & 0.1752 & 0.6898 
\end{tabular}
\caption{TREC 2021 Location (Non-Europeans as protected, Europeans as non-protected.)}
\end{subtable}

\caption{Pairwise evaluation results. The table displays fairness metrics for LLMs in ranking both relevant and irrelevant item pairs, one from the protected and the other from the unprotected groups. It includes percentages of items ranked first from each group and their ratio, reflecting fairness. The varying levels of fairness across LLMs, particularly in irrelevant pairings, highlight the importance of further enhancing fairness in LLMs.}
\label{tab_pairwise}
\end{table*}

\subsubsection{Item-side Analysis}
In Table~\ref{tab_listwise}, MonoT5 and MonoBERT exhibit robust Precision@20 scores, reflecting their effectiveness in ranking. However, their fairness metrics reveal a gap in equitable gender representation, with MonoT5 slightly outperforming MonoBERT on this front. This performance discrepancy is likely because these models utilize the complete text of Wikipedia pages, providing a wealth of features that represent the items more comprehensively. On the other hand, LLMs face constraints due to the maximum token limits for input, limiting their capacity to fully exploit the extensive textual information available in the TREC datasets, thereby impacting their ranking capability.

Among LLMs, including GPT-3.5, GPT-4, Mistral-7b, and Llama2-13b, the Precision@20 scores are comparatively lower than those of neural ranking models. This may reflect the generative models' broader focus beyond just ranking tasks.
The fairness metrics for these LLMs are varied. GPT-3.5 and GPT-4 manage to stay closer to the ideal fairness ratio, indicating a more balanced treatment of gender groups. Mistral-7b, while maintaining a similar precision, falls behind in fairness, indicating a potential gender bias in ranking. Llama2-13b, although consistent in its approach to fairness, reveals room for improvement in precision.

When contrasting neural rankers with LLMs, it becomes apparent that although neural rankers demonstrate higher precision, they do not necessarily outperform LLMs in terms of fairness. This observation underscores the importance of considering fairness, particularly for users who prioritize it over precision in specific applications. Within the LLM group, there is no uniformity in achieving fairness, suggesting that the models' training, design, and inherent biases may influence their ability to rank fairly.

\subsubsection{Query-side Analysis}
Analyzing the query-side fairness from the Table~\ref{tab_listwise}, our focus is on whether LLMs provide similar ranking performance for different query attributes (Male vs. Female, European vs. Non-European). It reveals a consistent trend across both neural ranking models and LLMs: they tend to favor female and European queries over male and Non-European ones. While fairness metrics for LLMs like GPT-3.5, GPT-4, Mistral-7b, and Llama2-13b are relatively close to 1, indicating an attempt at balanced treatment, the Precision@20 scores suggest a different story, with a clear skew towards female and European queries. This observed pattern, evident in both MonoT5 and MonoBERT, points to an underlying bias that persists despite efforts to achieve equitable treatment across query attributes, underscoring the need for enhanced model training and fairness optimization.

In Figure~\ref{fig_box_plot}, we plot the predicted ranking of the protected groups, highlights distinct patterns in fairness and ranking performance between neural rankers and LLMs. LLMs demonstrate tighter rank distributions but exhibit biases toward certain query attributes. For example, disparities are observed in the treatment of gender and geographic attributes, with both MonoT5 and MonoBERT often ranking female and non-European queries more favorably, a trend also noted to varying degrees within LLMs. This suggests that while neural rankers may excel in precision, LLMs offer more consistent rankings, though neither group is devoid of fairness issues. These findings emphasize the necessity for further tuning and bias mitigation in both neural rankers and LLMs to ensure equitable treatment across all query attributes.

\subsection{Pairwise Evaluation Results}

In the pairwise evaluations detailed in Table~\ref{tab_pairwise}, our focus is on assessing the fairness of various LLMs by studying how they rank pairs of items when both are considered relevant or irrelevant. The analysis aims to reveal whether these models display biases toward items from specific groups. GPT-3.5 consistently shows a preference for female items in both scenarios, with this inclination more pronounced for irrelevant items, suggesting a bias in favor of female items. Similarly, GPT-4 displays a moderate bias towards female items, with ratios indicating a stronger bias in irrelevant contexts. This observed trend across models and datasets signals an area for improvement, pointing to the need for more balanced algorithms that do not favor one group over another, particularly in situations where item relevance is neutral.

Contrastingly, Mistral-7b shows a distinct bias towards male items in relevant pairs, notably in the TREC 2022 Gender dataset, raising questions about the model's decision-making process and suggesting that its algorithm may weigh male items more heavily when they are relevant. However, this bias diminishes with irrelevant pairs, indicating a different algorithmic behavior in such contexts. Llama2-13b, on the other hand, presents a significant bias towards female items across all datasets, in both relevant and irrelevant pairs, which is concerning for its overall fairness. Overall, while some LLMs show nuanced biases, others like Llama2-13b require more interventions to ensure fair and equitable treatment across all group attributes.

\begin{figure*}[ht!]
\centering
\begin{subfigure}[b]{1\textwidth}
\includegraphics[width=\textwidth]{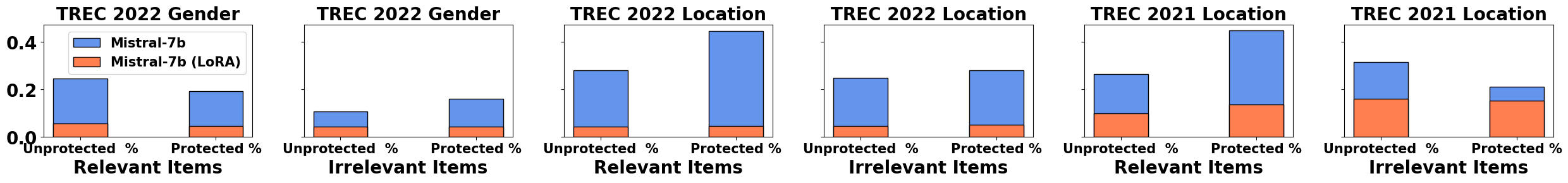}
   \caption{Percentage of protected vs. unprotected group items ranked first across different TREC datasets.}
\end{subfigure}

\begin{subfigure}[b]{1\textwidth}
   \includegraphics[width=\linewidth]{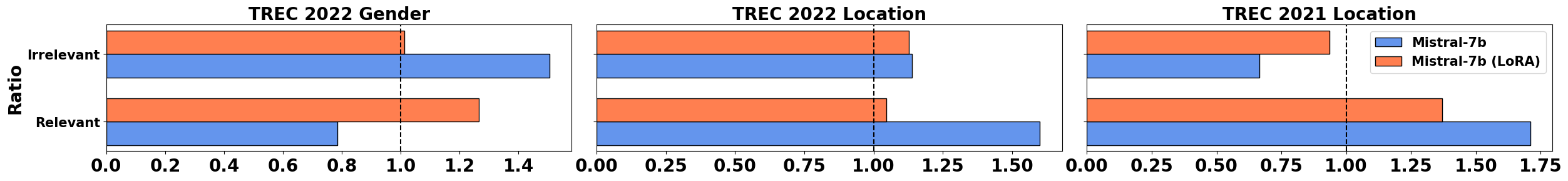}
   \caption{Ratio of protected over unprotected group across different TREC datasets.}
   \label{fig_lora_ratio}
\end{subfigure}
\caption{Impact of LoRA Fine-Tuning on Mistral-7b's Fairness. Figure (a) shows the percentage of first-ranked items from protected and unprotected groups, while Figure (b) demonstrates the resulting fairness ratios. The LoRA-adjusted model yields ratios closer to the ideal fairness benchmark of 1.0 across TREC datasets.}
\label{fig_lora}
\end{figure*}

\subsection{Overall Evaluation}
Overall, analyzing both the listwise and pairwise evaluation results in the Table~\ref{tab_listwise} and Table~\ref{tab_pairwise}, we observe a complex picture of fairness. While the listwise evaluation, based on group exposure ratios, suggests a fair representation of different groups, the pairwise evaluation reveals the unfairness in LLMs. This inconsistency is particularly evident when LLMs rank pairs of relevant and irrelevant items from protected and unprotected groups.

\section{Enhancing Fairness with LoRA}

We employed LoRA~\cite{hu2022lora} to fine-tune the Mistral-7b model. Our approach involves creating a balanced training dataset with equal representation of responses from both protected and unprotected groups. This balanced dataset aims to steer the model towards fairer rankings when evaluating pairs of relevant or irrelevant items from diverse groups. 
The implementation of the LoRA module is facilitated using the PEFT~\cite{peft} package. Aligning with the parameter-efficient methodology outlined in the original LoRA, our study specifically focuses on adapting attention weights. To simplify and enhance parameter-efficiency, we opted to freeze other parameters.
In our case, we set the optimal rank to 1, deeming a low-rank adaptation matrix as adequate. The chosen learning rate is 0.003, and the batch size is set at 4. These configurations were selected based on considerations specific to our study.
The dataset, comprising approximately 140,000 item pairs randomly sampled for each TREC dataset, facilitate comprehensive training. The process, conducted on an NVIDIA A100 80GB, needs approximately 30 hours. We split the queries for training and testing, using 80\% for training and the remaining 20\% for testing.

The results of fine-tuning Mistral-7b with LoRA are illustrated in Figure~\ref{fig_lora}. Post-tuning, there is a noticeable reduction in consistent responses from the model when queried twice with reversed item orders. This indicates an increase in response variability, which is a positive indicator of fairness, as less predictability in responses can mitigate systematic bias. The improvement in fairness is further supported by Figure~\ref{fig_lora_ratio}, where the outcomes post-LoRA fine-tuning show ratios approaching 1.0, indicating a more equitable treatment of protected and unprotected groups by the model.

\section{Conclusion}
The empirical study and in-depth analysis provided in this research study have revealed the intricate biases presented in Large Language Models (LLMs) when evaluated for fairness through listwise and pairwise methods. While listwise evaluations painted a picture of relative fairness, a deeper investigation via pairwise evaluations uncovered subtler and more profound biases that often favored certain protected groups. The implementation of LoRA fine-tuning on the Mistral-7b model yielded encouraging strides to rectify these biases, demonstrating enhanced fairness in the model's output. Going forward, our efforts will pivot towards further improving ranking performance with targeted ranking loss functions, while concurrently addressing fairness more holistically through refined prompting strategies, aiming for an optimal balance between utility and equity in broad LLM-based ranking applications. 

\section*{Ethics Statement}

In this research study, we empirically examine the fairness of LLMs when used as ranking algorithms (namely, LLM rankers). To conduct the proposed research, we mainly adopt publicly available datasets to test the ranking fairness of LLMs across a variety of contexts and demographic groups. We recognize that the use of LLM rankers has the potential to reflect and even exacerbate existing biases rooted in their training corpus. 

The objective of this work is not to advocate for or against the use of LLM rankers but to provide an empirical foundation upon which future discussions on the ethical use of LLMs can be built. We commit to presenting our findings in a manner that is objective and devoid of personal biases, with the hope that our work contributes to the responsible development of LLM technologies.

By acknowledging the complexities and responsibilities associated with our research, we aim to foster a deeper understanding of how LLMs can be used in ways that promote fairness and equity. We believe that through careful consideration and ethical diligence, the benefits of LLMs can be harnessed while mitigating their risks and ensuring they serve the interests of all segments of society.

\section*{Acknowledgements}
We would like to thank DOCOMO Innovations, Inc. for the support of this research.

\clearpage
\bibliography{custom}

\appendix

\end{document}